# Broken mirror symmetry tuned topological transport in PbTe/SnTe heterostructures

Feng Wei,[1,3] Chieh-Wen Liu,[2] Da Li,[1] Chun-Yang Wang,[1,3] Hong-Rui Zhang,[3,4] Ji-Rong Sun,[4] Xuan P. A. Gao,[2,*] Song Ma,[1,*] and Zhidong Zhang[1]

[1]Shenyang National Laboratory for Materials Science, Institute of Metal Research, Chinese Academy of Sciences, 72 Wenhua Road, Shenyang 110016, China

[2]Department of Physics, Case Western Reserve University, 2076 Adelbert Road, Cleveland, Ohio, 44106, United States

[3]University of Chinese Academy of Sciences, Beijing 100049, China

[4]Beijing National Laboratory for Condensed Matter Physics, Institute of Physics, Chinese Academy of Sciences, Beijing 100190, China

*email: xuan.gao@case.edu (X.P.A.G.), songma@imr.ac.cn (S.M.)

## Abstract

The tunability of topological surface states and controllable opening of the Dirac gap are of great importance to the application of topological materials. In topological crystalline insulators (TCIs), crystal symmetry and topology of electronic bands intertwine to create topological surface states and thus the Dirac gap can be modulated by symmetry breaking structural changes of lattice. By transport measurement on heterostructures composed of p-type topological crystalline insulator SnTe and n-type conventional semiconductor PbTe, here we show a giant linear magnetoresistance (up to 2150% under 14 T at 2 K) induced by the Dirac Fermions at the PbTe/SnTe interface. In contrast, PbTe/SnTe samples grown at elevated temperature exhibit a cubic-to-rhombohedral structural phase transition of SnTe lattice below 100 K and weak antilocalization effect. Such distinctive magneto-resistance behavior is attributed to the broken mirror symmetry and gapping of topological surface states. Our work provides a promising application for future magneto-electronics and spintronics based on TCI heterostructures.

Topological crystalline insulators (TCIs) are a new topological phase of matter whose topological invariant and conducting boundary states originate from the crystal lattice symmetry [1-4] instead of time-reversal symmetry of topological insulators (TIs) [5-8]. Unlike the conventional $Z_2$ TIs with layered structure [9], the TCIs have a rock salt crystal structure and possess even number of Dirac cones of topological surface states (TSSs) [3,4] resulting in the possibility of strong coupling with other materials through large proximity effects. Heterointerfacing TIs and band insulators (BIs) can induce diverse phases such as Dirac semimetal and Weyl semimetal [10-12]. Compared with TIs, the rich interplay between topology, crystal symmetry and electronic structure in TCIs [13,14] makes possible novel topological states in heterointerfaces of TCIs and BIs worth-while to study.

The narrow gap semiconductor SnTe with rock salt structure was the first theoretically predicted TCI and its TSS was expected to exist on highly symmetrical crystal surfaces such as {001}, {110} and {111} [2]. Subsequently, the presence of the TSSs protected by the mirror symmetry in SnTe was verified by angle-resolved photoemission spectroscopy (ARPES) experiments [1]. Motivated by the promising applications of TCIs in low power electronics and tunable spintronic devices [15], magnetotransport experiments probing the TSSs have been conducted. Weak antilocalization (WAL) effect [16-20] and two dimensional (2D) SdH oscillations [18] related to the TSSs have been detected in SnTe thin films. Aharonov-Bohm (AB) effect in single SnTe nanowires [19] and high field linear magnetoresistance (LMR) in In doped SnTe nanoplates [20] have been revealed. However, very little is known about the behavior of the TSSs with regard to lattice symmetry breaking and heterointerface formation. PbTe is a BI with rock salt structure and its lattice constant is close to SnTe, resulting in a lattice well matched interface when PbTe and SnTe form heterostructure. Previous studies demonstrated that PbTe/SnTe superlattice is a type Ⅱ superlattice exhibiting semimetallic state [21,22]. However, magnetotransport behavior and the effect of the TSS in the PbTe/SnTe heterointerface are still lacking, while the interplay between the electronic topology and crystal symmetry in PbTe/SnTe heterostructure may induce rich phenomena.

Here, we presented a systematic investigation on the magnetotransport behavior of the PbTe/SnTe heterostructures grown on $SrTiO_3(111)$ (STO) and demonstrated a striking change of magnetoresistance (MR) due to the structural change in SnTe lattice. For SnTe crystal, it is well known that there is a transition from cubic to rhombohedral lattice structure around 100 K and this transition shifts to lower temperatures at higher hole carrier densities [23-25]. Indeed, we observed no sign of structure phase transition in samples with high carrier densities. Moreover, PbTe/SnTe exhibits a large LMR similar to other gapless topological materials. On the other hand, when the carrier density is low and the SnTe lattice transforms into the rhombohedral phase, only WAL and a small classical parabolic MR is present. Structural and compositional analysis shows that the interfacial diffusion at the PbTe and SnTe interface plays an important role in controlling the carrier density and thus structural phase transition in SnTe. These results give a prominent example of using heterostructuring and mirror symmetry breaking to fundamentally tune the MR of the TCIs.

The epitaxial PbTe/SnTe heterostructures with bottom PbTe layer and top SnTe layer (structural schematic shown in Fig. 1(a)) were grown on SrTiO$_3$ (111) substrate by molecular beam epitaxy (MBE) (see methods for details in supplemental Material SA). The heterostructures sample numbers and corresponding film nominal thickness are shown in supplemental material SB Table 1, together with the samples' additional transport parameters. In Figure 1(b), reflection high-energy electron diffraction (RHEED) pattern of a heterostructure sample is shown. The sharp streaky RHEED pattern indicates high crystalline quality and atomically flat surface of the PbTe/SnTe heterostructure. The X-ray diffraction (XRD) pattern in Fig. 1(c) shows the (002) and (004) diffraction peaks of PbTe and SnTe, indicating that the growth direction of PbTe and SnTe is along [001] for these two layers. Fig. 1(d) shows the lattice image obtained by high resolution transmission electron microscopy (HRTEM) of a PbTe/SnTe sample grown at 200 ℃, indicating a good crystalline quality of our heterostructure. We found that the growth temperature critically affects the sharpness of the PbTe/SnTe interface. Fig. 1(e) presents the cross-sectional high angle annular dark field (HAADF) scanning transmission electron microscopy (STEM) image of a typical PbTe/SnTe hetero-structure grown at 200 ℃. The element mapping (for area in the red frame) is shown in Fig. 1(f). The sharp interface between PbTe and SnTe can be clearly seen in the elemental mapping. In comparison, the HAADF STEM and element mapping (red frame) of a sample grown at 260 ℃ reveals significant interlayer diffusion (Fig. 1(g) and (h)). As will be shown later, this inter-diffusion has a fundamental effect on the structure's transport characteristics and is essential for the cubic to rhombohedral phase transition and the breaking of mirror symmetry in this system.

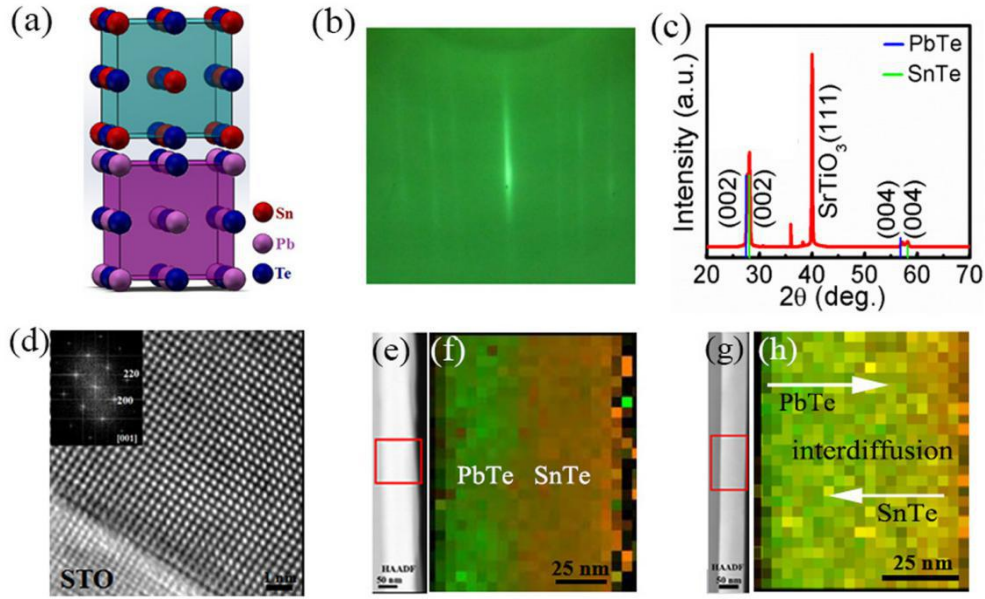

**Figure 1.** Structural characterization of PbTe/SnTe heterostructures. (a) The crystal structural schematic of PbTe/SnTe heterostructure. (b) RHEED pattern of a PbTe/SnTe heterostructure with sharp streak indicating high crystalline quality and atomically flat surface of the PbTe/SnTe heterostructure. (c) XRD data showing the growth of PbTe/SnTe thin film along the direction of [001] and meanwhile no other impurity phase is detected in present thin films. (d) High resolution transmission electron microscopy (HRTEM) of a PbTe/SnTe sample grown at 200 ℃ showing a good crystalline quality of our heterostructure which is consistent with the RHEED pattern. The Fourier transformation of Fig. 1d is shown in the inset, confirming the thin film's [001] orientation. (e, f) and (g, h) are HAADF STEM image and element mapping (mapping area marked by red frame) of a PbTe/SnTe sample grown at 200 ℃ and 260 ℃, respectively. Interdiffusion between Pb (green) and Sn (red) can be seen in (h) for sample grown at 260 ℃.

We have studied eleven samples with different nominal PbTe and SnTe thicknesses. These samples were grown either at 200 ℃ or 260 ℃. Table 1 gives a list of all the eleven samples studied in this work and their respective parameters (nominal PbTe, SnTe thicknesses, sheet resistance $R_s$ at 300K and 2K etc.). The samples are numbered according to the order of their growth and measurements. After examining all the samples' transport behavior, we find that their transport characteristics fall into two categories, correlated with the sample growth temperature and aforementioned interface quality/interdiffusion. Specifically, at low temperatures (e.g. 2 K), the low temperature sheet resistance $R_s$ of sample #1, 2, 4, 7, 8, 9 is about three orders of magnitude lower than that of sample #3, 5, 6, 10, 11, although all the samples have comparable sheet resistances at room temperature.

This striking difference is further illustrated in Fig. 2(a) which shows the temperature dependent sheet resistance of all the eleven samples. While all the samples have similar $R_s$ at 300 K, the temperature dependence of $R_s$ shows a two to three orders of magnitude drop below ~100 K for sample #1, 2, 4, 7, 8, 9, giving rise to their low $R_s$ at low $T$. On the other hand, sample #3, 5, 6, 10, 11 show a much weaker metallic behavior below 100 K, leading to a much higher $R_s$ at low $T$. The distinct transport property between these two types of samples is also exhibited in the magneto-resistances. Fig. 2(b) presents the longitudinal magneto-resistance (MR) defined as $[R_s(B)-R_s(B=0)]/R_s(B=0)$ of sample #1 which has a sharp PbTe/SnTe interface and low $R_s$ at 2 K. It can be seen that an appreciable MR starts to appear at $T < 50$ K and the MR grows larger and takes the linear dependence on the magnetic field as $T$ deceases. At 2 K, the MR reaches 2150% under 14 T field. The other samples (#2, 4, 7, 8, 9) grown at 200 ℃ and with sharp PbTe/SnTe interface also exhibit similar MR behavior (Supplementary Fig. S1).

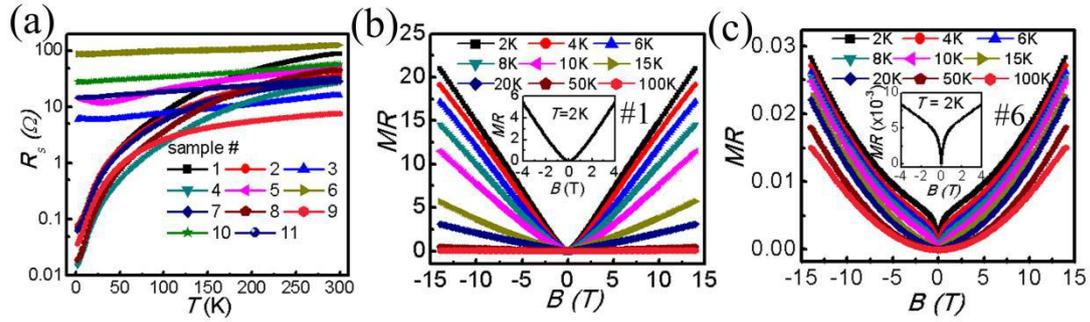

**Figure 2.** Two types of temperature and magnetic field dependent transport in PbTe/SnTe. (a) Temperature-dependent sheet resistance (in logarithmic scale) of PbTe/SnTe heterostructures in zero magnetic field. (b) Magneto-resistance (MR) defined as $[R_s(B)-R_s(B=0)]/R_s(B=0)$ at various temperatures $T = 2$-100 K for sample #1. A giant linear magnetoresistance appears at low temperatures (larger than 2000% under 14 T at 2 K). (c) MR at $T = 2$-100 K for sample #6 where a small (up to ~3% under 14 T at 2 K) quadratic-like behavior is seen together with a weak antilocalization induced dip around $B=0$ at low temperatures. Inset: zoomed in view of the low field MR data at $T = 2$ K.

On the other hand, samples with significant interlayer diffusion exhibited MR behavior at low $T$ distinct from the ones with a sharp interface. Figure 2(c) presents the MR of sample #6. Comparing to the giant linear MR of sample #1, the low temperature MR of sample #6 is much smaller (< 3% under 14 T) and consists of a dip below $B\sim2$ T and a parabolic MR at higher $B$. The MR dip around $B = 0$ T becomes sharper at lower $T$, a signature of the weak antilocalization (WAL) effect. WAL is a quantum correction effect to the classical Drude conductivity due to the destructive interference of wavefunctions in materials with strong spin orbit coupling [26]. In the 2D diffusive transport, the WAL quantum correction to the magneto-conductance can be fitted by the standard Hikami-Larkin-Nagaoka (HLN) equation [27] with the fitting parameters of α and $l_\Phi$ where 2α and $l_\Phi$ denotes the number of conduction channels and phase coherent length, respectively. Because SnTe has four Dirac cones on the (001) plane, α can range from 0.5 to 4 depending on whether there is strong coupling between the TSSs and bulk states or its mirror symmetry is broken. The magneto-conductance of sample #6 was well described by the HLN equation with fitting parameters α = 0.46 and $l_\Phi$ = 128 nm at $T$ = 2 K (Supplementary Fig. S2), indicating a single surface transport channel. Given the 2D quantum coherence transport in sample #6, we believe a Dirac cone is on the surface of sample #6.

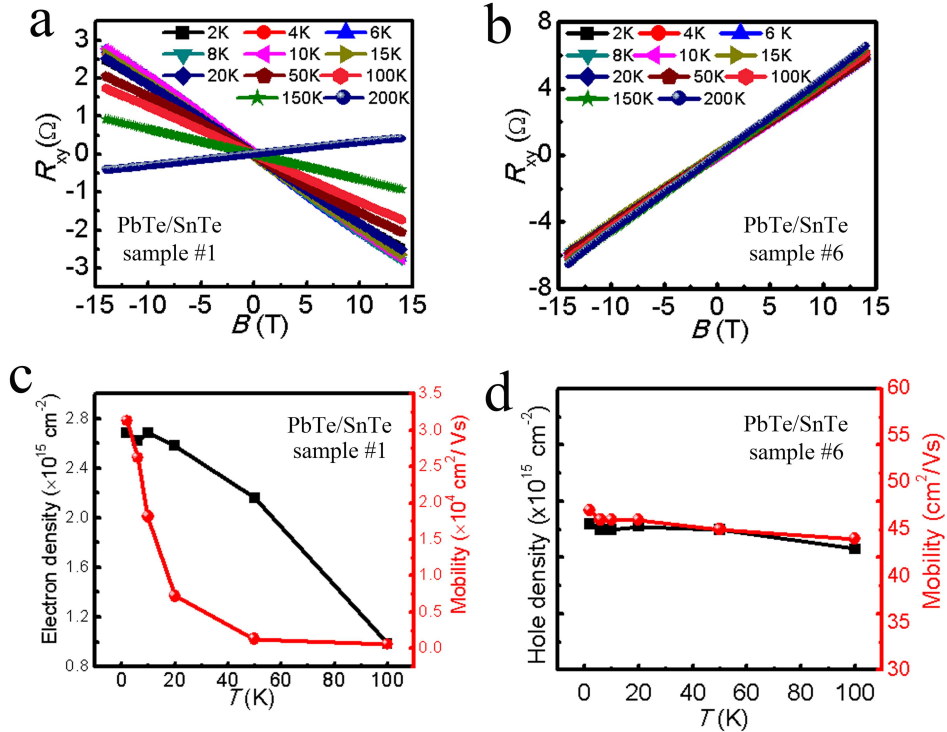

**Figure 3.** Temperature-dependent Hall resistance $R_{xy}$ and carrier density/mobility. (a, b) $R_{xy}$ vs. $B$ of sample #1 and #6. While sample #6 shows p-type behavior for the whole temperature range, sample #1 exhibits p type to n type transition at $T$ < 200 K. (c, d) Temperature dependent carrier density and mobility of sample #1 and #6 extracted by fitting the magneto-conductance data.

In addition, there are two clear differences in the Hall resistance effect between these two types of samples showing either large LMR or weak MR at low *T*. Fig. 3(a) and (b) present the Hall resistance data $R_{xy}(B)$ at various temperatures for sample #1 and #6 (Hall data for other samples are shown in Supplementary Fig. S3). Firstly, in samples showing large LMR at low *T* (represented by sample #1), there is a crossover from p-type conduction to n-type conduction around 150-200 K, whereas the samples showing weak MR at low T (represented by sample #6) exhibit p-type conduction in the whole temperature range investigated. Secondly, we also note that sample #1 has much smaller Hall slope than sample #6 in the p-type conduction regime at high temperatures (T > 150 K), indicating much higher hole density in sample #1 in the high *T* regime. Similar contrasting behavior were exhibited by other samples showing large LMR in comparison with the ones showing much weaker MR at low *T*. Table 1 lists the extracted hole density at 150 K or 200 K when all the samples are in the high *T* p-conduction regime. It is seen that samples with sharp PbTe/SnTe interface (and giant low temperature LMR) generally have hole densities on the order of $10^{16}$-$10^{17}$/cm$^2$, an order of magnitude higher than the samples with interface diffusion (and weak low temperature MR). Quantitative Hall analysis (see Supplementary Fig. S4) revealed that the strong metallic drop in sample #1 is associated with a sharp increase in the mobility of n-type carriers below 100K, reaching ~3×10$^4$ cm$^2$/Vs at 2 K (Fig. 3(c)). On the other hand, sample #6's hole mobility is much lower, ~ 45 cm$^2$/Vs, and nearly temperature independent down to 2K (Fig. 3(d)).

Nonsaturating LMR observed in various novel material systems has attracted long lasting research interest. Early experiments on metals with open Fermi surface (e.g. Bi [28]) showed magneto-resistance increasing linearly with the magnetic field at high fields. More recently, LMR has been observed in numerous materials with gapless electronic states (e.g. graphene [29], topological insulators [30,31], Dirac and Weyl semimetals [32,33]). While there have been many theoretical models proposed to explain the LMR and no consensus is reached yet, the two most discussed models are Parish and Littlewood (PL)'s disorder induced mobility fluctuation model [34,35] and Abrikosov's quantum LMR model [36,37]. In the mobility fluctuation model, current in a strongly disordered medium at large fields may flow perpendicular to the applied voltage, contributing the linear field dependent Hall resistance $\rho_{xy} \propto B$ to the longitudinal magnetoresistance [34,35]. However, the observation of LMR in PbTe/SnTe heterostructures with better interface crystalline quality and much higher mobility contradicts the disorder related origin of the LMR. Abrikosov's quantum LMR model points out that LMR can appear in materials with the gapless linear energy spectrum in extreme quantum limit where all the carriers occupy only the lowest Landau level (LL) [36,37]. Although the quantum LMR model has the attractive aspect of relating to gapless topological materials with linear dispersion, it also has challenges in explaining the persistence of LMR down to low fields where multiple LLs are filled and the quantum limit condition is not satisfied. In any case, regardless the specific mechanism for the LMR, the LMR observed in PbTe/SnTe heterostructures with high mobility and sharp interfaces suggests that it is related to the gapless interface/surface states in TCI SnTe. Actually, the PbTe/SnTe heterostructure is of type II heterostructure [21,22], where the valence band edge of SnTe is higher than the conduction band edge of PbTe by ~0.3 eV at 77K [21]. Given the very high hole carrier density (~$10^{16}$/cm$^2$ or $10^{21}$-$10^{22}$/cm$^3$) in PbTe/SnTe samples grown at low temperature (200 ºC), the Fermi energy in SnTe is deep in the valence band (calculated to be 0.36eV below valence band, using hole density $10^{21}$/cm$^3$ and mass $m^*=m_e$, the free electron mass) and at a lower position than the Fermi energy in n-type PbTe. Thus the equilibration of Fermi level between PbTe and SnTe leads to electron transfer from the n-type PbTe to p-type SnTe and a downward band-bending is expected for the conduction and valence bands of SnTe at the interface as shown in Fig. 4(a). This downward shift in the energy bands of SnTe also shifts the Dirac point accordingly and causes the Fermi level above the Dirac point at the interface, giving rise to n-type Dirac electrons. Given the low mobility of electrons in bulk PbTe and holes in bulk SnTe, we attribute the high mobility electron carriers dominating the low *T* transport and LMR to n-type Dirac electrons at the PbTe/SnTe interface where the Dirac point of TCI SnTe's surface states is pulled to a position lower than the bulk valence band edge (Fig. 4(a)).

In PbTe/SnTe samples with lower hole density and interface diffusion, two effects that critically affect the electronic properties of the sample are expected. First, it has been established that a cubic-to-rhombohedral structure phase transition occurs in SnTe with low hole density below 100 K, and the transition temperature decreases with increasing hole density. Eventually, the transition temperature drops to zero at hole density > ~ $10^{21}$/cm$^3$ [23] and SnTe maintains the cubic structure down to the lowest temperature. Using 20 nm thick SnTe as an example, this critical hole density is ~ $2\times10^{15}$/cm$^2$. It can be seen from Table 1 that most samples without LMR have hole densities lower than this critical density and the cubic-to-rhombohedral structure phase transition is expected. However, all the samples exhibiting LMR and sharp interface have hole densities much higher than the critical density and the cubic lattice structure of SnTe is expected to be intact down to the lowest temperature. Indeed, we observed the signature of the cubic-to-rhombohedral structure phase transition in SnTe in samples #3, 5, 6, 10, 11. We replot the *R* vs. *T* curves of samples #3, 5, 6, 10, 11 in logarithmic scale to show the trend of these samples more clearly in Fig. S5a. As highlighted by the vertical arrows, a small hump is observed at *T* <~ 100 K in these samples in Fig. S5a. Correspondingly, the significant difference is further highlighted by dR/dT vs. T in Fig. S5b where the shade purple box denotes phase transition starting point temperature $T_{up}$ and ending temperature $T_{down}$, which is consistent with the feature associated with the cubic-to-rhombohedral structure phase transition in previous studies of SnTe [23-25].

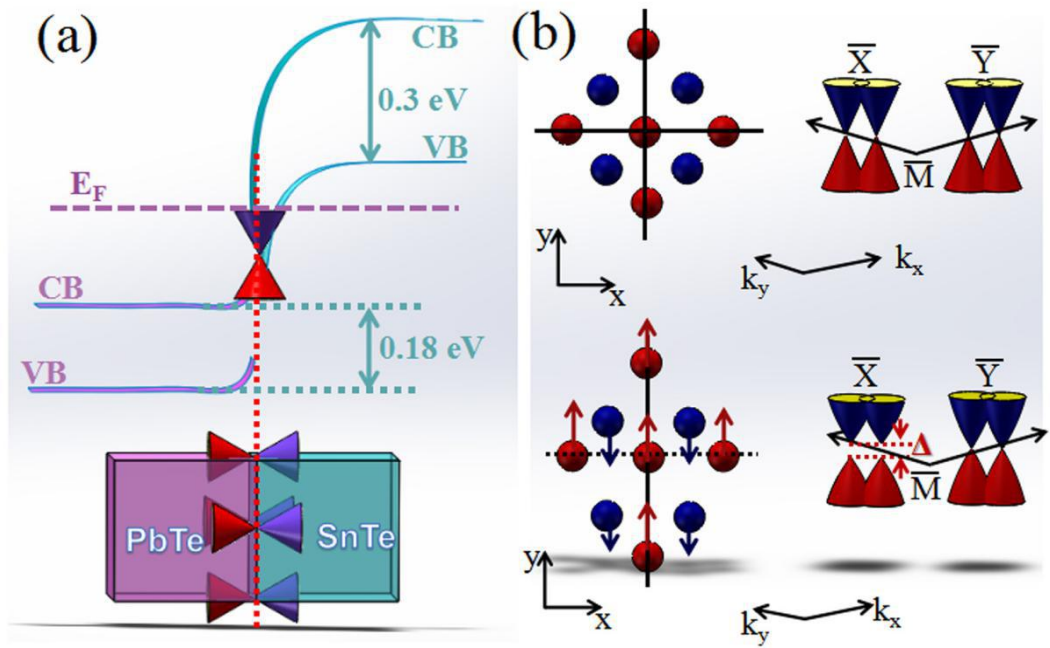

**Figure 4** Schematic energy band diagram for PbTe/SnTe heterostructures and structural phase transition induced mirror symmetry breaking in SnTe. (a) The type II heterointerface and broken gap alignment in PbTe/SnTe with sharp interface. The band bending at the interface also lowers the Dirac point. (b) Schematic mirror symmetry breaking resulting in the opening of an energy gap Δ at a Dirac point due to the cubic-rhombohedral structural phase transition on the surface of SnTe. The surface mirror line is indicated by solid lines. The dash line indicates the mirror symmetry breaking due to the distorted SnTe (001) surface. The arrows correspond to the displacement direction of Sn atom (red) and Te atom (blue).

A significant impact of the cubic-to-rhombohedral structure phase transition in PbTe/SnTe samples with low hole density is the breaking of lattice mirror symmetry as shown the schematic in Fig. 4(b). Theoretically, such lattice distortion induced broken mirror symmetry is anticipated to gap out some or all of the Dirac points in the topological surface states of SnTe and as such the linear dispersion relation is also broken [14]. The absence of LMR and high mobility carrier transport in samples with interface diffusion and low hole density is thus most likely due to the broken mirror symmetry induced destruction of the Dirac topological surface state.

In conclusion, we have systematically investigated the magneto-transport properties of PbTe/SnTe heterostructures grown at different substrate temperatures that affect the interface quality. We find that PbTe/SnTe heterostructures with sharp interface and high hole densities display p-to-n transition and a giant LMR at low temperatures related to the gapless TSS with high mobility at the interface between TCI SnTe and band insulator PbTe. On the other hand, in PbTe/SnTe structures with significant interface diffusion and low hole densities, low mobility p-type conduction is observed throughout the whole temperature range and only weak MR is exhibited at low temperatures. The presence of cubic-to-rhombohedral structure phase transition in the latter category of PbTe/SnTe structures strongly suggest the important role of broken lattice mirror symmetry in the disappearance of giant LMR and high mobility TSSs transport in these samples.

This work was supported by the Natural Science Foundation of China (NSFC) with the Nos. 51571195, 51331006, and 51590883 and 11520101002. We also gratefully acknowledge support from the National Key R&D Program of China (No. 2017YFA0206301). C.W.L. and X.P.A.G. thank NSF (grant number DMR-1607631) for supporting the work at CWRU. We thank Wan-rong Geng for kind helps in TEM sample preparation.

# Supplemental material for

**Broken mirror symmetry tuned topological transport in PbTe/SnTe heterostructures**

Feng Wei, Chieh-Wen Liu, Da Li, Chun-Yang Wang, Hong-Rui Zhang, Ji-Rong Sun, Xuan P. A. Gao, Song Ma, and Zhidong Zhang

## SA.  Experimental Methods

**MBE growth.** The growth of PbTe/SnTe heterostructure was performed using ultrahigh vacuum MBE system with base pressure below $2\times10^{-10}$ mbar. Insulating $SrTiO_3(111)$ (STO) substrates were cleaned by ultrasonic in deionized water at 75 ℃ for 100 min. Then the $SrTiO_3(111)$ (STO) substrates were outgassed at 600 ℃ for 120 min before the growth of films. High-purity Pb(99.999%), Sn(99.999%) and Te(99.9999%) were evaporated from Knudsen effusion cells and their evaporation temperature is 500 ℃, 980 ℃ and 300 ℃, respectively. To grow PbTe/SnTe samples with sharp interface (sample #1, 2, 4, 7, 8, 9). PbTe thin film was first deposited on the STO substrate which was maintained at 200 ℃. Then the as-prepared PbTe thin film was annealed at 200 ℃ for 90 min. Next, SnTe thin film was deposited on the PbTe thin film which was maintained at 200 ℃. Subsequently, the SnTe thin film was annealed at 200 ℃. For comparison, in sample #3, 5, 6, 10, and 11, the PbTe and SnTe thin films were deposited on STO substrate at higher temperature.

The PbTe thin film was deposited at 270 °C and subsequently annealed at 270 °C for 90 min. And the SnTe thin film was deposited on PbTe thin film at 260 °C and annealed at 260 °C for 90 min.

**Structural characterization.** The reflection high-energy electron diffraction (RHEED) pattern of PbTe/SnTe heterostructure was collected from in situ RHEED in MBE chamber. The X-ray diffraction (XRD, Bruker, D8 diffractometer, Cu Kα radiation) and high-resolution transmission electron microscopy (HRTEM, FEI Tecnai F20) was utilized to characterize the crystal structure of PbTe/SnTe heterostructures. The cross-sectional high angle annular dark field (HAADF) scanning transmission electron microscopy (STEM) image and element mapping of a typical PbTe/SnTe hetero-structure were collected by the STEM (FEI Tecnai F20).

**Transport measurement.** The transport measurements were performed *ex situ* on the PbTe-SnTe heterostructure. To avoid possible contamination, a 3-nm-thick epitaxial Te capping was deposited on the PbTe-SnTe heterostructure at room temperature before removal from the growth chamber of MBE. The Hall effect and longitudinal resistance were measured in a Quantum Design Physical Property Measurement System (PPMS; 2K, 14 T) by a standard six-probe method on rectangular samples on which the contacts were made with indium near the perimeter.

## SB. Table for the experimental samples

**Table 1.** The various transport parameters ($R_s$, carrier density and mobility etc.) for different nominal thickness of PbTe and SnTe layer of PbTe/SnTe heterostructures. ($R_s$: sheet resistance).

| sample number | PbTe | SnTe | $R_s$ (Ω) at 300 K | $R_s$ (Ω) at 2 K | Carrier density at 150 K (/cm$^2$) | Mobility at 150 K (cm$^2$/V s) |
|---|---|---|---|---|---|---|
| #1 | 20 nm | 5 nm | 89.61 | 0.069 | 2.02×10$^{16}$ (200 K) | 230 (200 K) |
| #2 | 20 nm | 20 nm | 36.29 | 0.079 | 9.42×10$^{15}$ | 48.4 |
| #4 | 5 nm | 20 nm | 28.33 | 0.015 | 7.53×10$^{16}$ | 12.9 |
| #7 | 20 nm | 10 nm | 29.14 | 0.064 | 3.26×10$^{17}$ | 1.5 |
| #8 | 5 nm | 5 nm | 45.36 | 0.019 | 1.52×10$^{16}$ | 37.9 |
| #9 | 20 nm | 40 nm | 7.67 | 0.037 | 1.59×10$^{16}$ | 91.8 |
| #3 | 20 nm | 40 nm | 16.22 | 5.90 | 2.10×10$^{15}$ | 323.9 |
| #5 | 40 nm | 20 nm | 50.84 | 14.57 | 1.08×10$^{15}$ | 209.0 |
| #6 | 5 nm | 5 nm | 126.46 | 88.16 | 1.38×10$^{15}$ | 43.9 |
| #10 | 20 nm | 10 nm | 58.70 | 28.30 | 8.16×10$^{14}$ | 199.2 |
| #11 | 40 nm | 20 nm | 32.76 | 14.57 | 1.26×10$^{15}$ | 248.7 |

**Fig. S1** Magneto-resistance (MR) at various temperatures $T$ = 2-100K for sample #2, 4, 7, 8, 9 which is defined as $[R_s(B)-R_s(B=0)]/R_s(B=0)$.

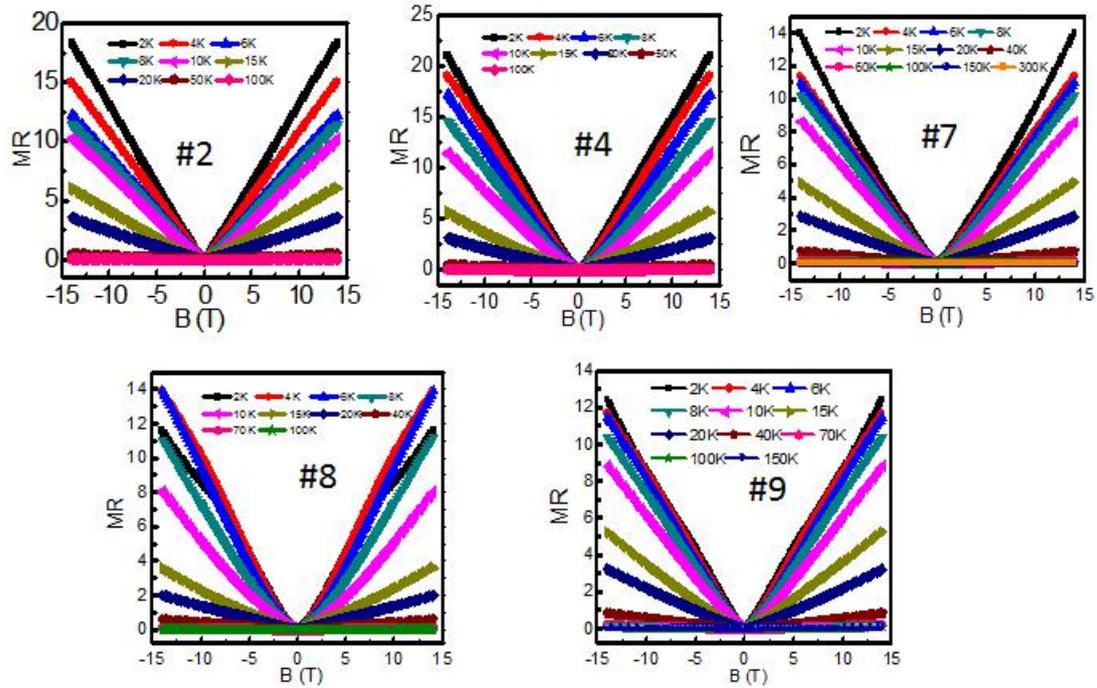

**Fig. S2** Magnetoconductance analysis for #6. **a,** The schematic for the magnetotransport measurements of the PbTe/SnTe heterostructure. **b,** The magnetic field dependence of magnetoconductance under perpendicular magnetic fields ΔG(B, 0º) (dark yellow line) and that under parallel magnetic fields ΔG(B, 90º) (magenta line) for #6, where ΔG(B, θ) = G(B, θ) - G(0, θ) = 1/R(B, θ) - 1/R(0, θ). **c,** The temperature dependent subtracted magnetoconductance ΔG(B, 0º) - ΔG(B, 90º) and the red line stands for the fitting using HLN formula.

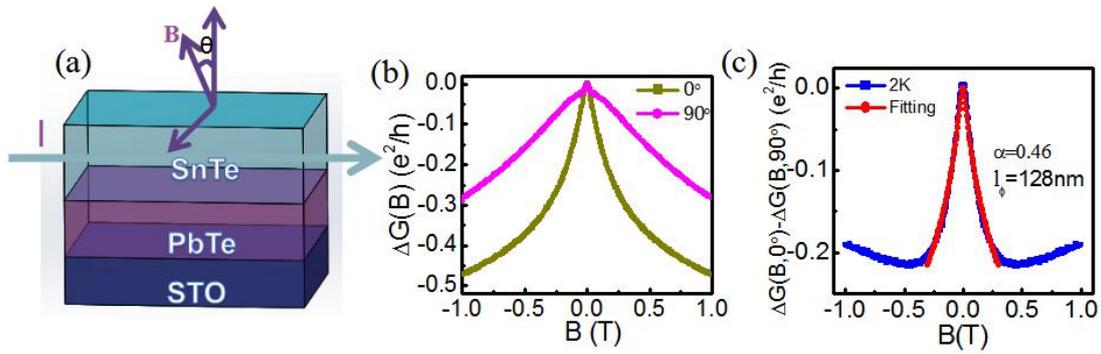

In the transport of the TSSs, due to the spin-momentum locking for the Dirac electrons traveling along a self-intersecting path, a shift of π in the Berry phase gives rise to the destruction interference of the wavefunction, resulting in an enhancement of the conductivity[S1]. When a magnetic field perpendicular to the closed path is applied, the effect of interference will fall off and a reduction of the conductivity will arise as shown Fig. S2b. According to the theory of quantum correction in the 2D diffusive transport, the negative MC due to the WAL effect can be given by the HLN equation [S2]:

$$\Delta\sigma_{xx}^{2D} = \sigma_{xx}^{2D}(B_\perp) - \sigma_{xx}^{2D}(0) = -\alpha\frac{e^2}{2\pi^2\hbar}\left[\ln\left(\frac{\hbar}{4eB_\perp l_\varphi^2}\right) - \Psi\left(\frac{1}{2} + \frac{\hbar}{4eB_\perp l_\varphi^2}\right)\right]$$

Here $\sigma_{xx}$ indicates the longitude component of conductivity and the superscript 2D indicates that the equation is valid for a two-dimensional conducting sheet. α represents a coefficient related to the number of coherent transport channels. e is the electronic charge. $\hbar$ is Planck's constant divided by 2π. B is the magnetic field perpendicular to the 2D plane. $l_\Phi$ is the phase coherent length. As shown in Fig. S2c, we use the HLN equation to fit the 2D negative MC and extract the fitting parameter α and $l_\Phi$.

**Fig. S3** Temperature dependent Hall resistance $R_{xy}$ for sample #2, 3, 4, 5, 7, 8, 9, 10, 11. The $R_{xy}$ shows a crossover from p-type conduction to n-type conduction for sample #2, 4, 7, 8, 9 whose MR exhibit nonsaturating LMR. The $R_{xy}$ shows p-type conduction in the whole temperature range for sample #3, 5, 10, 11 whose MR exhibit WAL.

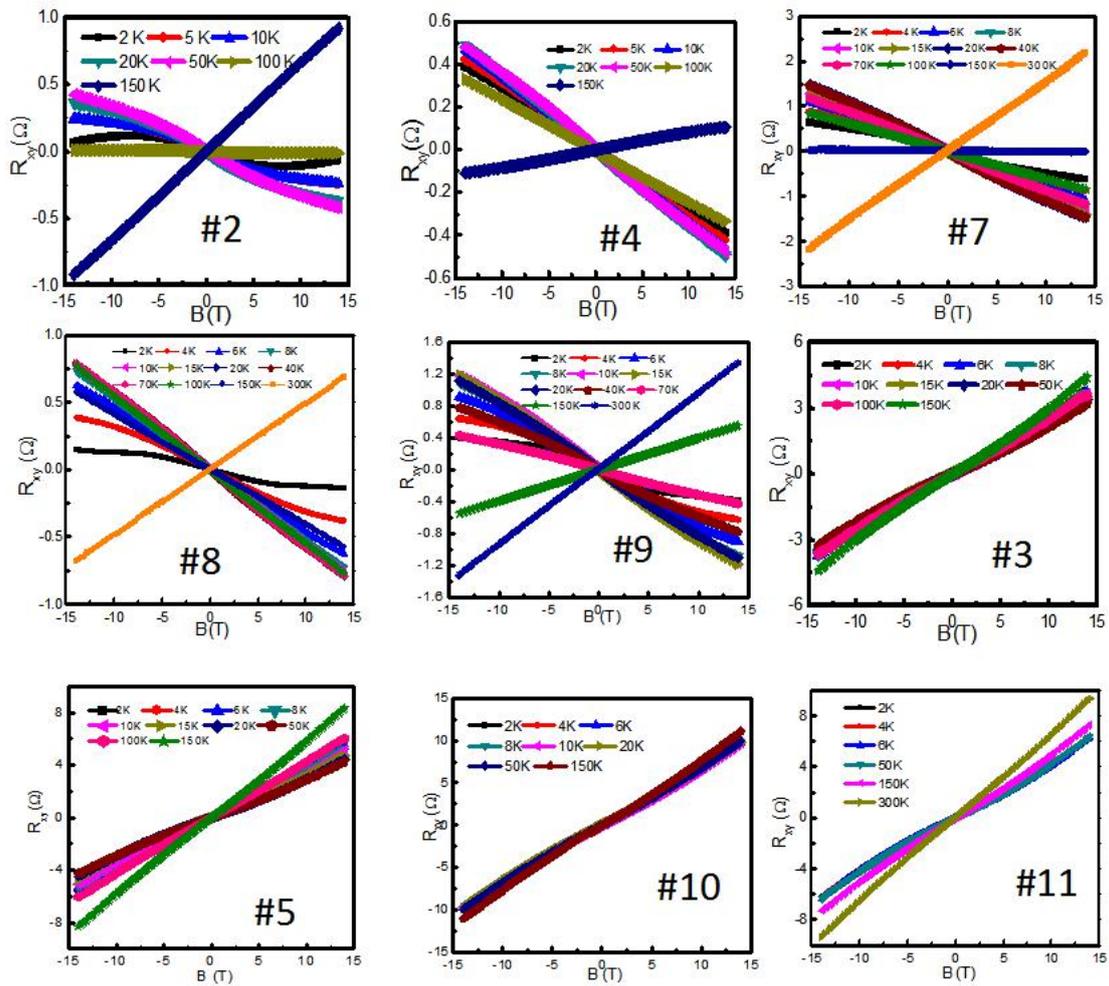

**Fig. S4.** Magnetotransport properties analysis. **a,** Hall conductance data at 2 K of sample #1 fitted with the two-band Drude model. **b,** Hall resistance data at 2 K of sample #6 fitted with the one-band Drude model.

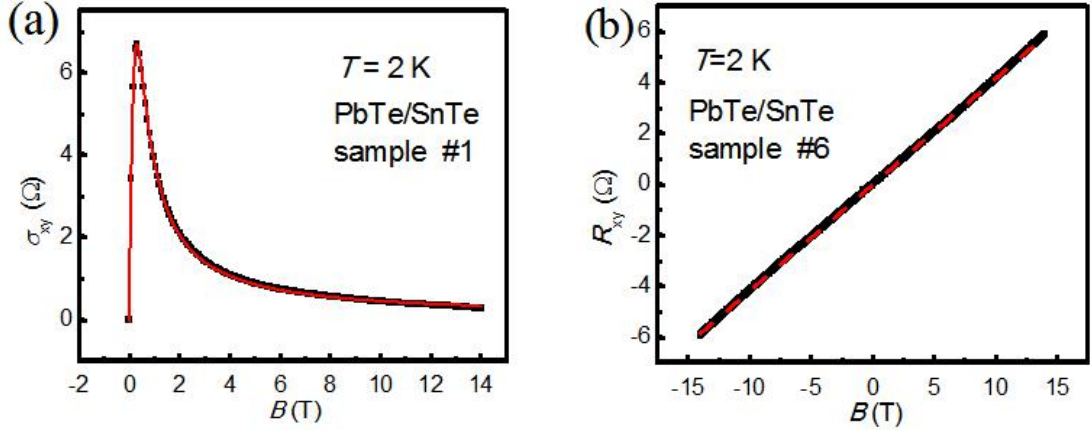

The temperature dependence of carrier density and mobility of sample 1 (Fig. 3 c) is obtained by fitting the magnetoconductivity data to the two-band Drude model which is given by

$$\sigma_{xx} = eB\left(\frac{n_1\mu_1}{1+\mu_1^2 B^2} + \frac{n_2\mu_2}{1+\mu_2^2 B^2}\right); \sigma_{xy} = eB\left(\frac{n_1\mu_1^2}{1+\mu_1^2 B^2} + \frac{n_2\mu_2^2}{1+\mu_2^2 B^2}\right),$$

where $n_1$ and $n_2$ are carrier densities of two different types of carriers with mobilities $\mu_1$ and $\mu_2$, respectively. To reduce the number of fitting parameters, the limit $B\to 0$ is applied to $\sigma_{xx}(B)$ and $\sigma_{xy}(B)$ such that

$$\frac{\sigma_{xx}(0)}{e} = n_1\mu_1 + n_2\mu_2 = C_1; \lim_{B\to 0}\frac{\sigma_{xy}}{eB} = n_1\mu_1^2 + n_2\mu_2^2 = C_2,$$

where $C_1$ and $C_2$ are constants that can be determined directly from the measured data. Therefore, the two parameters $n_1$ and $n_2$ can be eliminated and the Hall conductance equation can be rewritten as

$$\sigma_{xy}(B) = eB \left[ \frac{C_1\mu_1 - C_2}{\left(\frac{\mu_1}{\mu_2} - 1\right)(1+\mu_2^2 B^2)} + \frac{C_1\mu_2 - C_2}{\left(\frac{\mu_2}{\mu_1} - 1\right)(1+\mu_1^2 B^2)} \right],$$

where the mobilities can be obtained and the corresponding carrier densities are then calculated[S3].

Since resistance is the quantity that is directly measured, the Hall conductivity data in Fig. S 4a is obtained by inverting the resistivity tensor:

$$\sigma_{xy} = \frac{\rho_{xy}}{\rho_{xx}^2 + \rho_{xy}^2}$$

The red curve is the fit to the Hall conductivity data, which yields $n_1 = 2.69 \times 10^{15}$ /cm$^2$, $\mu_1 = 31270 \pm 40$ cm$^2$/Vs, $n_2 = 2.60 \times 10^{17}$ /cm$^2$ and $\mu_2 = 24 \pm 4$ cm$^2$/Vs.

To obtain the transport properties in Fig. 3d, we fit the data to the one-band Drude model since the Hall data in sample #6 shows a linear dependence. Fig. S4b shows the Hall resistance as a function of magnetic field at 2K. The red dashed curve represents the fit to the following equation based on the one-band Drude model

$$R_{xy} = \frac{\mu B}{\sigma_{xx}(0)},$$

where the fitted mobility and the carrier density $n$ can be calculated from $n = \frac{\sigma_{xx}(0)}{\mu e}$. The fitted mobility at 2K is 47 cm$^2$/Vs and the corresponding hole carrier density is $1.52 \times 10^{15}$/cm$^2$.

**Fig. S5** The transport evidence for the cubic to rhombohedral structure phase transition. **a,** The *R* vs. *T* curves of samples #3, 5, 6, 10, 11 with low hole density in logarithmic scale. A small hump is highlighted by the vertical arrow in every plot which indicates the cubic-to-rhombohedral structure phase transition[S4]. **b,** dR/dT vs. T for the corresponding samples in **a**. The shade purple box denotes phase transition starting point temperature $T_{up}$ and ending temperature $T_{down}$.

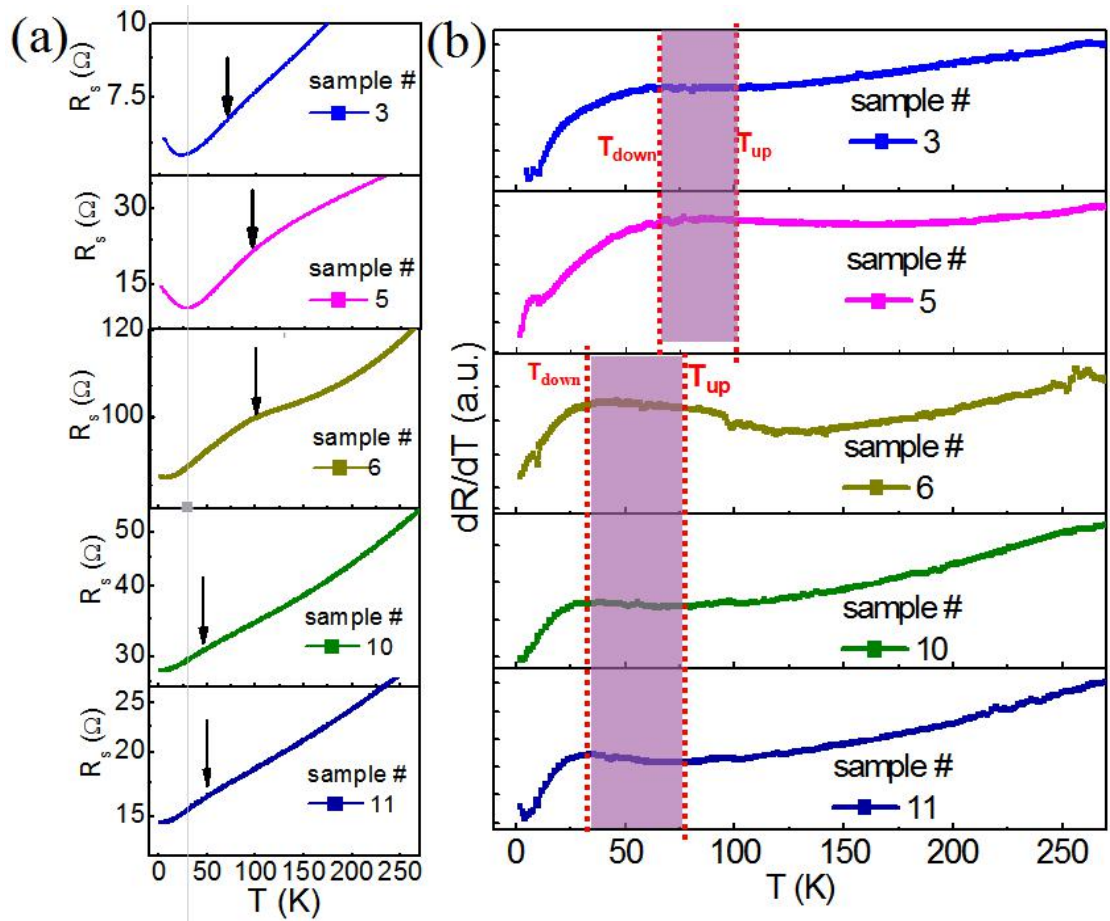